\documentclass[preprint,showpacs,preprintnumbers,amsmath,amssymb]{revtex4}
\usepackage{amsfonts}

\usepackage{graphicx}
\usepackage{dcolumn}

\begin{document}

\title{All-Optical Control of Light Group Velocity with a Cavity Optomechanical System }

\author{Wei He}
\author{Jin-Jin Li}
\author{Ka-Di Zhu}
\email{zhukadi@sjtu.edu.cn}
\affiliation{%
Department of Physics, Shanghai Jiao Tong University, 800 Dong Chuan
Road, Shanghai 200240, China
}%
\date{\today}

\begin{abstract}
We theoretically demonstrate complete all-optical control of light
group velocity via a cavity optomechanical system composed of an
optical cavity and a mechanical resonator. The realization depends
on no specific materials inside the cavity, and the control of light
group velocity stems from the interaction between the signal light
and the moving optical diffraction grating within the cavity in
analogy to the stimulated Brillouin scattering(SBS). Furthermore, we
show that a tunable switch from slow light to fast light can be
achieved only by simply adjusting the pump-cavity detuning. The
scheme proposed here will open a novel way to control light velocity
by all-optical methods in optomechanical systems.
\end{abstract}

\maketitle

\section{Introduction}
The ability to control the velocity of light has attracted a lot of
attention from both technical and scientific communities in physics
\cite{1}. Devices based on slow and fast light can be used as
buffering and regeneration in optical telecommunication,
continuously tunable phase shifter in microwave photonics and
spectrometer in interferometry \cite{2,3,4,5}. Rapid progress of
slow and fast light has been made in a variety of media by several
physical mechanisms, such as electromagnetically induced
transparency (EIT), coherent population oscillation (CPO) and
stimulated Brillouin scattering (SBS) \cite{6,7,8,9,10}. Recently,
light control in solids has been observed in crystals, erbium doped
fibers and photo-refractive materials \cite{11,12,13}. Moreover,
extreme values of group velocity at room temperature, which are
suitable for many practical applications, have also been reported
\cite{14}. On the other hand, bestriding the realms of classical and
quantum mechanics, optomechanical system has been subject to
increasing investigation\cite{15,16}. By coupling a driven
high-frequency mode to a high-Q, low-frequency mechanical mode and
with the analysis of the intrinsic properties, such system offers
great promise for a huge variety of applications and fundamental
researches\cite{17,18,19,20,21}.

In the present article, we propose a novel all-optical scheme to
control light group velocity by using a cavity optomechanical
system. Theoretical analysis shows that when the cavity-pump
detuning is fixed properly, the cavity optomechanical system,
composed of an optical cavity and a mechanical resonator, displays a
strong dispersive property with little absorption of signal light,
and thus the control of group velocity can be realized. The physical
picture behind is, to some extent, similar to the simulated
Brillouin scattering which occurs readily in any transparent
material. However, the process that the signal light experiences in
the system is a complete all-optical process, since no specific
material is needed to be placed in the cavity. And, unlike the EIT
process, fast light can also be generated in the optomechanical
system. Moreover, we find that the scattering of signal light
depends on the pump-cavity detuning, hence the switch between slow
and fast light can be achieved by adjusting pump-cavity detuning
properly.

\section{Theory}
We consider the canonical situation in which a driven high-finesse
cavity is coupled by momentum transfer of the cavity photons to a
micromechanical resonator.  The physical realization as shown in
Fig.1(a) is a Fabry-Perot cavity being formed on one end by a moving
mirror. The cavity free spectrum range $L/2c$ ($c$ is the speed of
light in the vacuum and $L$ is the effective cavity length
\cite{22,23}) is much larger than the frequency of movable mirror
($\omega_{m}$). Therefore the scattering of photons to other cavity
modes can be ignored, and we can adopt the single-cavity-mode
description. Before we discuss the system's ability to control the
group velocity, we start our description with the familiar
Hamiltonian in the rotating frame at a pump field frequency
$\omega_{pu}$ \cite{24,25}
\begin{eqnarray}
H=&&\hbar \Delta_{pu} a^{+}a+\frac{1}{2} \hbar
\omega_{m}(p^{2}+q^{2}) - \hbar G_{0}a^{+}a q + i \hbar E_{pu}
(a^{+}-a)+i\hbar E_{s}(a^{+}e^{-i\delta t}-ae^{i\delta t}),
\end{eqnarray}
where $\Delta_{pu}=\omega_{c}-\omega_{pu}$ is the detuning of the
pump field frequency and the frequency $\omega_{c}$ of the cavity
mode with bosonic operators $a$ and $a^{+}$. The quadratures $q$ and
$p$ associate to the mechanical mode satisfying the usual
commutation relations of canonical coordinates. The parameter
$G_{0}=\frac{\omega_{c}}{L}\sqrt{\frac{\hbar}{m\omega_{m}}}$ is the
coupling rate between the cavity and the resonator ($m$ is the
effective mass of mechanical mode \cite{23}). $E_{pu}$ and $E_{s}$
are related to the laser power $P$ by
$\left\vert{E_{pu}}\right\vert=\sqrt{\frac{2P_{pu}\kappa}{\hbar\omega_{pu}}}$
and
$\left\vert{E_{s}}\right\vert=\sqrt{\frac{2P_{s}\kappa}{\hbar\omega_{s}}}$
respectively ($\kappa$ is the cavity amplitude decay rate).
$\delta=\omega_{s}-\omega_{pu}$ is the detuning of signal and pump
field, $\omega_{s}$ is the signal field frequency.

According to the Heisenberg equation of motion
$i\hbar\frac{dO}{dt}=[O,H]$, the temporal evolutions of the lowering
operator $a$ and the dimensionless position operator $q$ are given
by
\begin{eqnarray}
\frac{da}{dt}=-i\Delta_{pu}a +iG_{0}aq+E_{pu}+E_{s}e^{-i\delta t},
\end{eqnarray}
\begin{eqnarray}
\frac{d^{2}q}{dt^{2}}+\omega^{2}_{m}q=\omega_{m}G_{0}a^{+}a.
\end{eqnarray}

In what follows we ignore the quantum properties of $a$ and $q$
\cite{27,28,29}. Also let $\langle{a}\rangle$,
$\langle{a^{+}}\rangle$ and $\langle{q}\rangle$ be the expectation
values of operators $a$, $a^{+}$, and $q$. Thus by adding the
corresponding noise terms, the semiclassical equations for $a$ and
$q$ will be
\begin{eqnarray}
\frac{d\langle{a}\rangle}{dt}=-(i\Delta_{pu}+\kappa)\langle{a}\rangle
+iG_{0}\langle{a}\rangle\langle{q}\rangle+E_{pu}+E_{s}e^{-i\delta
t},
\end{eqnarray}
\begin{eqnarray}
\frac{d^{2}\langle{q}\rangle}{dt^{2}}+\gamma_{m}\frac{d\langle{q}\rangle}{dt}
+\omega_{m}^{2}\langle{q}\rangle=\omega_{m}G_{0}\langle{a^{+}}\rangle\langle{a}\rangle,
\end{eqnarray}
where $\gamma_{m}$ is the damping rate of mechanical mode. In order
to solve these equations, we make the ansatz \cite{30}:
$\langle{a(t)}\rangle=a_{0}+a_{+}e^{-i\delta t}+a_{-}e^{i\delta t}$
and $\langle{q(t)}\rangle=q_{0}+q_{+}e^{-i\delta t}+q_{-}e^{i\delta
t}$. Upon substituting these equations into equations (4) and (5),
and upon working to the lowest order in $E_{s}$ but to all orders in
$E_{pu}$, we obtain in the steady state
\begin{eqnarray}
a_{+}=E_{s}[\frac{-i\delta+(-i\Delta_{pu}+\kappa)+C}{(\kappa-i\delta)^{2}+(\Delta_{pu}+iC)^{2}-D}],
\end{eqnarray}
where $A=\frac{G_{0}^{2}}{\omega_{m}^{2}}$,
$B=\frac{\omega_{m}^{2}}{\omega_{m}^{2}-i\gamma_{m}\delta-\delta^{2}}$,
$C=iA\omega_{m}w_{o}+iAB\omega_{m}w_{o}$,
$D=A^{2}B^{2}\omega_{m}^{2}w_{0}^{2}$ and $w_{0}=|a_{0}|^{2}$. Here
parameter $w_{0}$ is determined by the equation:
\begin{eqnarray}
w_{0}[\kappa^{2}+(\Delta_{pu}-\frac{G_{0}^{2}}{\omega_{m}}w_{0})^{2}]=E_{pu}^{2}.
\end{eqnarray}

In order to investigate the dispersion and absorption property of
the system, we need to calculate the output field by using the
input-output relation $a_{out}(t)+a_{in}(t)=\sqrt{2\kappa}a(t)$
\cite{31}, where $a_{out}(t)$ is the output operator and $a_{in}(t)$
is the input operator with zero mean value. In accordance with the
above discussions, we also ignore the quantum properties of
$a_{out}(t)$ and $a_{in}(t)$ , and thus we can obtain
\begin{eqnarray}
\langle{a_{out}(t)}\rangle=a_{out0}+a_{out+}e^{-i\delta
t}+a_{out-}e^{i\delta t}=\sqrt{2\kappa}(a_{0}+a_{+}e^{-i\delta
t}+a_{-}e^{i\delta t}).
\end{eqnarray}
From this equation, we see that $a_{out+}$ equals to $\sqrt{2\kappa
}a_{+}$, which is a parameter in analogy to the linear optical
susceptibility. The real part of $a_{out+}$ exhibits absorptive
behavior, and its imaginary part shows dispersive property, for the
reason that the phase of light changes $\dfrac{\pi}{2}$ on the
reflection.

In such a system, the transmission of the probe beam is given by
\begin{eqnarray}
t_{p}(\omega_{s})=\dfrac{a_{out}}{a_{in}}=1-a_{out+}=\dfrac{P^2-Q^2-E_{s}\sqrt{2\kappa}(MP-NQ)}{P^2-Q^2}+i\dfrac{\sqrt{2\kappa}(NP+MQ)}{P^2-Q^2},
\end{eqnarray}
where $M=\kappa-G\gamma_{m}\delta E$, $N=G(F+1)-\delta-\Delta_{pu}$,
$P=\kappa^2-\delta^2+\Delta_{pu}^2+G^2(2F+1)-2\Delta_{pu}G(F+1)$,
$Q=-2\kappa\delta+2G^2\gamma_{m}^2\delta
E-2\Delta_{pu}G\gamma_{m}\delta E$,
$E=\dfrac{\omega_{m}^2}{(\omega_{m}^2-\delta^2)^2+\gamma_{m}^2\delta^2}$,
$F=E(\omega_{m}^2-\delta^2)$ and $G=A\omega_{m}w_{0}$. Then the
phase of signal light can be written as follows
\begin{eqnarray}
\phi(\omega_{s})=arg(t_{p}(\omega_{s}))=arg(\dfrac{E_{s}\sqrt{2\kappa}(NP+MQ)}{P^2-Q^2-E_{s}\sqrt{2\kappa}(MP-NQ)}).
\end{eqnarray}
This rapid phase dispersion can lead to a group delay $\tau_g$ given
by
\begin{eqnarray*}
\tau_{g}=-\dfrac{d\phi}{d\omega_{s}}=E_{s}\sqrt{2\kappa}(\dfrac{(NP+MQ)[2PP'-2QQ'-E_{s}\sqrt{2\kappa}(M'P+MP'-N'Q-NQ')]}{[P^2-Q^2-E_{s}\sqrt{2\kappa}(MP-NQ)]^2+2E_{s}^2\kappa(NP+MQ)^2}
\end{eqnarray*}
\begin{eqnarray}
-\dfrac{(N'P+NP'+M'Q+MQ')[P^2-Q^2-E_{s}\sqrt{2\kappa}(MP-NQ)]}{[P^2-Q^2-E_{s}\sqrt{2\kappa}(MP-NQ)]^2+2E_{s}^2\kappa(NP+MQ)^2}),
\end{eqnarray}
where $M'=-G\gamma_{m}(E+\delta E')$, $N'=GF'-1$, $P'=-2\delta+2G^2
F'-2\Delta_{pu}GF'$, $Q'=-2\kappa+2G^2\gamma_{m}^2(E+\delta
E')-2\Delta_{pu}G\gamma_{m}(E+\delta E')$,
$E'=-\dfrac{E^2}{\omega_{m}^2}[4(\omega_{m}^2-\delta^2)\delta+2\gamma_{m}^2\delta]$
and $F'=E'\omega_{m}^2-(E'\delta^2+2\delta E)$. Obviously, this
analytic expression for the induced time delay is very complicated,
so we have to calculate it numerically (see Fig.3 below).

\section{Results and Discussions}
Before proceeding, we note that such an optomechanical system
contains an optical property which is in analogous to the
electrostriction or optical absorption in real material systems. As
shown in Fig.1(a), when the pump field turns on, the circulating
light illuminates on and gives rise to a radiation pressure force
that deflects the mirror. In turn, the change of cavity's length
alters the distribution of circulating intensity. This variation
acts as an all optical diffraction grating in the cavity field
moving back and forth with the oscillation frequency $\omega_{m}$ of
the mechanical resonator. While the signal light travels in the
cavity, the mutual interaction between input lights and the grating
leads to the scattering of photons. If the signal light moves in the
same direction as the diffraction grating, pump photons will be
scattered into signal light, and hence a Stokes process occurs. On
the contrary, if they move in different directions, signal light
will be scattered into pump field, which results in an anti Stokes
process. Such a behavior is very similar to the stimulated Brillouin
scattering (SBS) in real material systems, in which an acoustic wave
of frequency $\Omega$ is produced by the mutual interaction between
light fields and material system. Through the process of
electrostriction, the material system responses to the input fields
by the fluctuations of dielectric constant which act as a moving
diffraction grating with frequency $\Omega$ as shown in Fig.1(b)
\cite{1,30}. In view of Kramers-Kronig relations, the cavity
optomechanical system will display a strong dispersive property
while gain or loss resonance occurs, and therefore the control of
light group velocity can be achieved.

Then, to prove this basic idea, we choose a realistic optomechanical
system \cite{32} to illustrate the numerical results. Fig.2 plots
both the real part and the imaginary part of $a_{out+}$ as a
function of signal-cavity detuning with $\Delta_{pu}=\mp10MHz$
respectively. Other parameters used in calculation are $E_{pu}=2MHz,
\kappa=2\pi\times215KHz$, $\omega_{m}=10MHz$ and
$\gamma_{m}=2\pi\times140Hz$\cite{32}. It is clear that the
dispersion curves (Fig.2(b) and Fig.2(d)) are very steep around the
center, which leads to the variation of light group velocity. At the
same time the absorption spectrum (Fig.2(a) and Fig.2(c)) splits
into two peaks (Normal Mode Splitting) at $\Delta_{s}=0$ which
ensures that the signal light passes through with little energy
loss. In Fig.3(a) and Fig.3(b) we plot $\tau_{g}$ as a function of
the amplitude of pump field $E_{pu}$ with $\Delta_{pu}=\mp10MHz$
respectively. While cavity-pump detuning is fixed at $-10MHz$, the
slope of the dispersion curve is positive and a slow-light process
occurs. As we can see, the group velocity of signal light is very
sensitive to $E_{pu}$. In Fig.3(a), the slope becomes steeper as
$E_{pu}$ decreases, leading to an increasingly lower group velocity.
Similarly, if the cavity-pump detuning shifts from $-10MHz$ to
$10MHz$, the signal pulse will experience a fast-light process.
Therefore it is possible to realize the switch between slow and
superluminal light by simply choosing a proper pump-cavity detuning.
The physical origin of these results is due to the interaction
between signal light and cavity field, as discussed above. The
fluctuation of light intensity inside cavity acts as a moving
diffraction grating composed of large amount of photons, and gives
rise to a large contribution to the phase dispersion. When the
signal light passes through, the scattering of photons occurs which
changes the transmitting time of the signal light.

We also use another model to discuss the property of the
optomechanical system. Although the cavity is empty, we regard it as
a material system composed of photons. As usual, we determine the
group velocity of light as
$v_{g}=\frac{c}{n+\omega_{s}(dn/d\omega_{s})}$ \cite{33,34}, where
the refractive index $n\approx1+2\pi\chi_{eff}$, and then
$\frac{c}{v_{g}}=1+2\pi
Re[\chi_{eff}(\omega_{s})]_{\omega_{s}=\omega_{c}}+2\pi \omega_{s}
Re(\frac{d\chi_{eff}}{d\omega_{s}})_{\omega_{s}=\omega_{c}}.$ Here
$\chi_{eff}$ is the effective susceptibility and is in direct
proportion to $a_{out+}$. Noticing that the phase of light has
changed on the reflection and
$Im[\chi_{eff}(\omega_{s})]_{\omega_{s}=\omega_{c}}=0$, the group
velocity index should be written as $n_g=\frac{c}{v_{g}}\approx
2\pi\omega_{c}Im(\frac{d\chi_{eff}}{d\omega_{s}})_{\omega_{s}=\omega_{c}}\propto
Im(\frac{da_{out+}}{d\omega_{s}})_{\omega_{s}=\omega_{c}}$. In
Fig.3(c) and Fig.3(d), we plot the group velocity index as a
function of the amplitude of pump field. The results are similar to
that of Fig.3(a) and Fig.3(b).

\section{Conclusions}
In conclusion, we have presented the slow light and the superluminal
light in a cavity optomechanical system which is composed of an
optical cavity and a mechanical resonator by a fully all-optical
method. The control of light group velocity  is achieved with no
specific materials placed inside the cavity. Also the switching
between slow and fast light can be realized easily by simply
adjusting the cavity-pump detuning. Finally, we hope that our scheme
proposed here can be realized by experiment in the near future.

\vskip 2pc \leftline{\bf Acknowledgement}

The part of this work has been supported by  National Natural
Science Foundation of China (No.10774101 and No.10974133), the
National Ministry of Education Program for Ph.D. and Key Laboratory
of Artificial Structures and Quantum Control, Ministry of Education.

\newpage
\centerline{\large{\bf Figure Captions}}

Fig.1 (a) Schematic diagram of a Fabry-Perot cavity with a movable
mirror in the presence of a strong pump field and a weak signal
field. (b) Schematic diagram of the SBS process in a real material
system. $\omega_{pu}$ and $\omega_{s}$ are the frequencies of pump
field and signal field respectively, and $\Omega$ is the frequency
of acoustic wave in the material system.

Fig.2 Plot of both the real part and the imaginary part of
$a_{out+}$ with $\Delta_{pu}=\mp10MHz$ respectively. Other parameter
values are $E_{pu}=2MHz$, $\omega_{m}=10MHz$,
$\kappa=2\pi\times215KHz$, $\gamma_{m}=2\pi\times140Hz$ and
$G_{0}=1.2MHz$.

Fig.3 (a) and (b) are group delay
$\tau_{g}=-\dfrac{d\phi}{d\omega_{s}}$ of signal light as a function
of the amplitude of driving field with $\Delta_{pu}=\mp10MHz$
respectively. (c) and (d) are the group velocity index
$n_{g}=\dfrac{c}{v_{g}}$ of slow light versus $E_{pu}$ with
$\Delta_{pu}=\mp10MHz$ respectively. Other parameter values are
$\omega_{m}=10MHz$, $\kappa=2\pi\times215KHz$,
$\gamma_{m}=2\pi\times140Hz$ and $G_{0}=1.2MHz$.

\clearpage
\begin{figure}
\includegraphics[width=15cm]{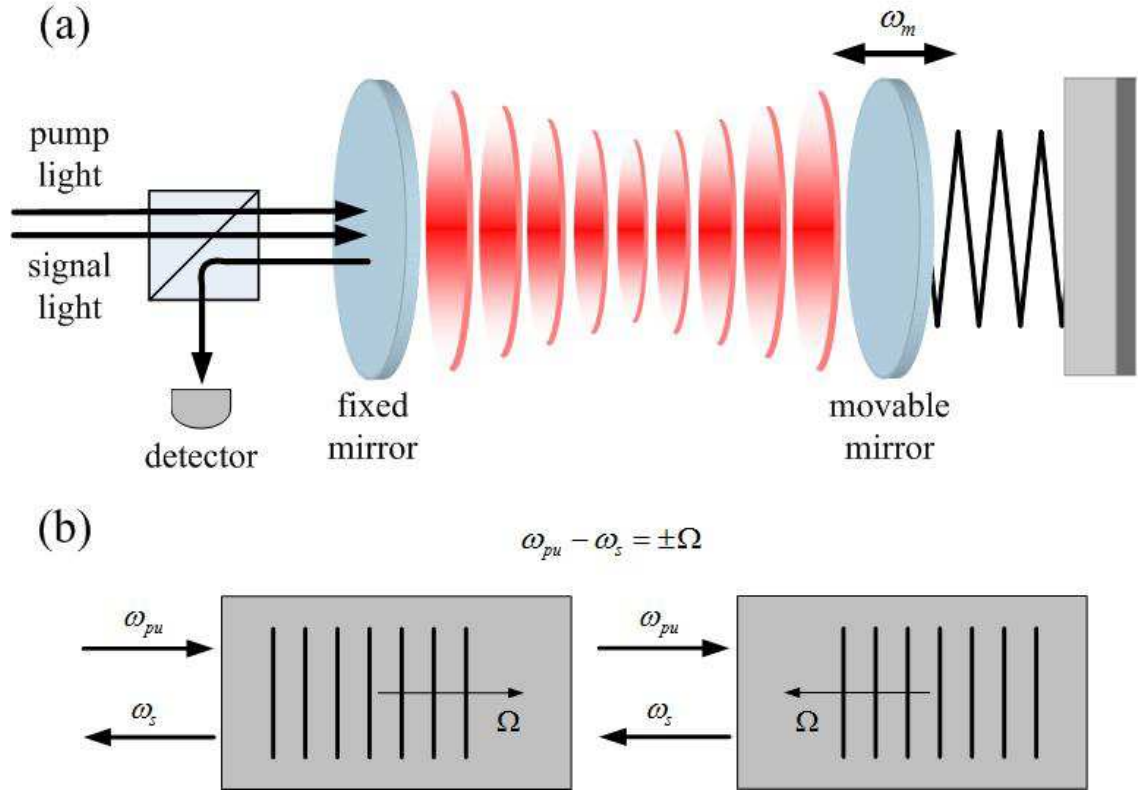}
\caption{(a) Schematic diagram of a Fabry-Perot cavity with a
movable mirror in the presence of a strong pump field and a weak
signal field. (b) Schematic diagram of the SBS process in a real
material system. $\omega_{pu}$ and $\omega_{s}$ are the frequencies
of pump field and signal field respectively, and $\Omega$ is the
frequency of acoustic wave in the material system.}
\end{figure}

\clearpage
\begin{figure}
\includegraphics[width=15cm]{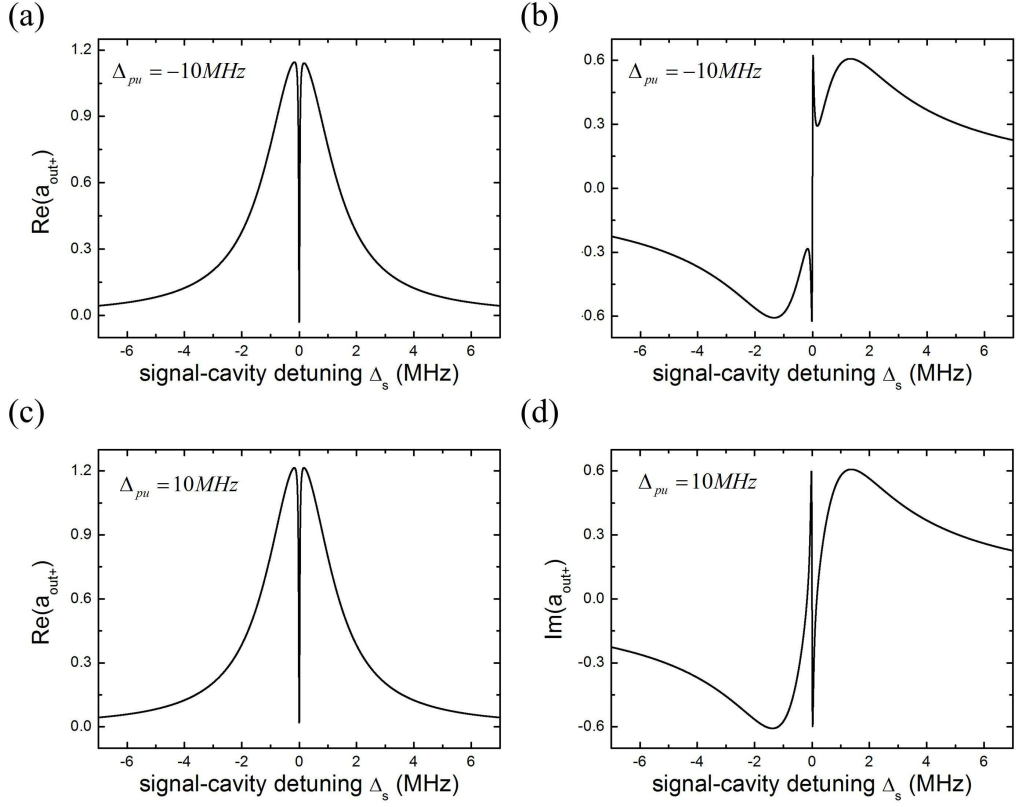}
\caption{Plot of both the real part and the imaginary part of
$a_{out+}$ with $\Delta_{pu}=\mp10MHz$ respectively. Other parameter
values are $E_{pu}=2MHz$, $\omega_{m}=10MHz$,
$\kappa=2\pi\times215KHz$, $\gamma_{m}=2\pi\times140Hz$ and
$G_{0}=1.2MHz$.}
\end{figure}

\clearpage
\begin{figure}
\includegraphics[width=15cm]{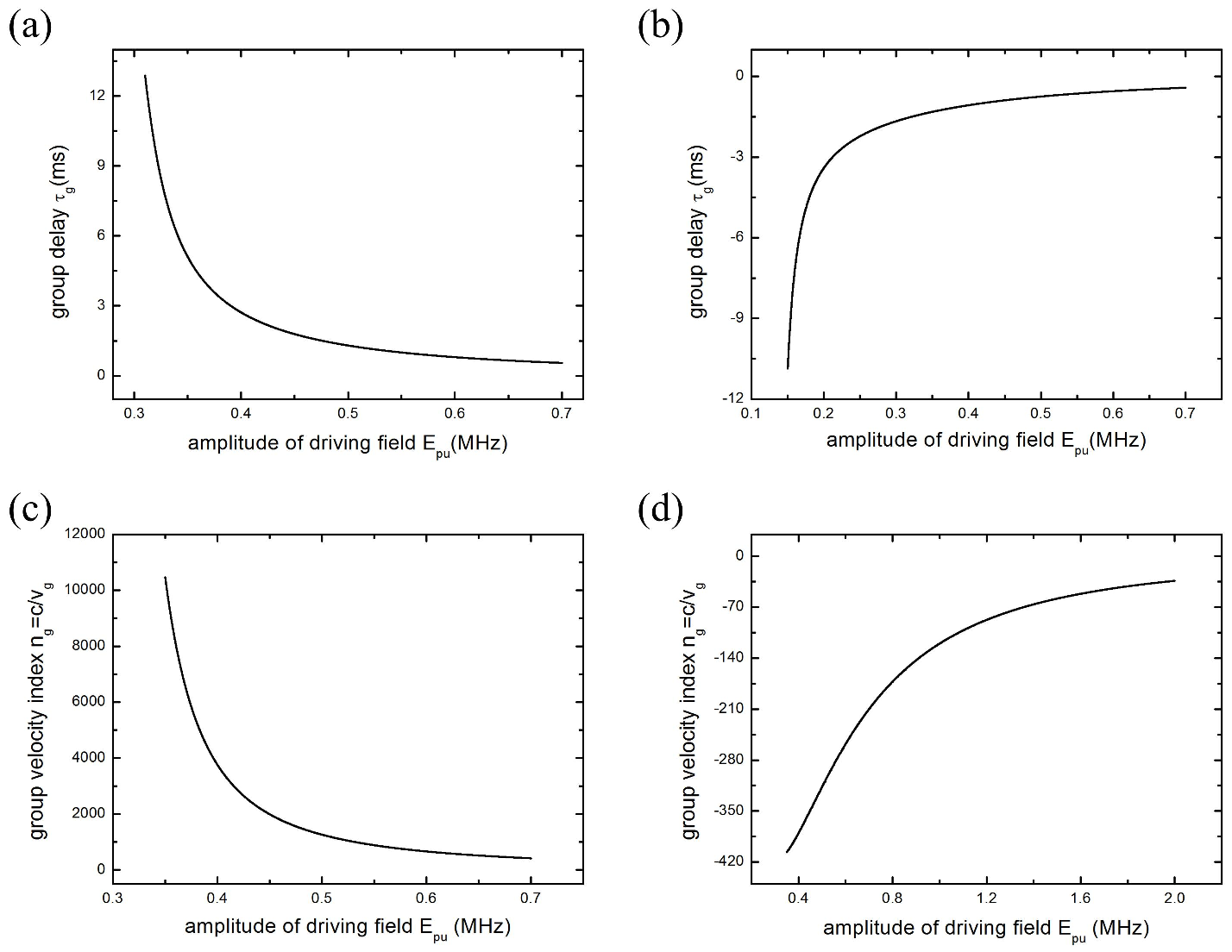}
\caption{(a) and (b) are group delay
$\tau_{g}=-\dfrac{d\phi}{d\omega_{s}}$ of signal light as a function
of the amplitude of driving field with $\Delta_{pu}=\mp10MHz$
respectively. (c) and (d) are the group velocity index
$n_{g}=\dfrac{c}{v_{g}}$ of slow light versus $E_{pu}$ with
$\Delta_{pu}=\mp10MHz$ respectively. Other parameter values are
$\omega_{m}=10MHz$, $\kappa=2\pi\times215KHz$,
$\gamma_{m}=2\pi\times140Hz$ and $G_{0}=1.2MHz$}
\end{figure}


\begin{thebibliography}{85}
\expandafter\ifx\csname
natexlab\endcsname\relax\def\natexlab#1{#1}\fi
\expandafter\ifx\csname bibnamefont\endcsname\relax
  \def\bibnamefont#1{#1}\fi
\expandafter\ifx\csname bibfnamefont\endcsname\relax
  \def\bibfnamefont#1{#1}\fi
\expandafter\ifx\csname citenamefont\endcsname\relax
  \def\citenamefont#1{#1}\fi
\expandafter\ifx\csname url\endcsname\relax
   \def\url#1{\texttt{#1}}\fi
\expandafter\ifx\csname urlprefix\endcsname\relax\def\urlprefix{URL
}\fi \providecommand{\bibinfo}[2]{#2}
\providecommand{\eprint}[2][]{\url{#2}}

\bibitem{1} R. W. Boyd and D. J. Gauthier, Science 326, 5956 (2009).
\bibitem{2} D. Dahan and G. Eisenstein, Opt. Express 13, 6234-6249 (2005).
\bibitem{3} R. W. Boyd, D. J. Gauthier and A. L. Gaeta, Optics and Photonics News 19, 18-23 (2006).
\bibitem{4} Z. Shi, R. W. Boyd, D. J. Gauthier and C. C. Dudley, Opt. Lett. 32, 915 (2007).
\bibitem{5} B. Macke and B. S¨¦gard, Phys. Rev. A 78, 013817 (2008).
\bibitem{6} M. M. Kash, V. A. Sautenkov, A. S. Zibrov, L. Hollberg, G. R. Welch,
M. D. Lukin, Y. Rostovtsev, E. S. Fry and M. O. Scully, Phys. Rev.
Lett. 82, 5229 (1999).
\bibitem{7} M. S. Bigelow, N. N. Lepeshkin, and R. W. Boyd, Phys. Rev. Lett. 90, 113903 (2003).
\bibitem{8} M. Fleischhauer, A. Imamoglu and J. P. Marangos, Rev. Mod. Phys. 77, 633-673 (2005).
\bibitem{9} Y. Okawachi, M. S. Bigelow, J. E. Sharping, Z. M. Zhu, A.
Schweinsberg, D. J. Gauthier, R. W. Boyd and A. L. Gaeta, Phys. Rev.
Lett. 94, 153902 (2005).
\bibitem{10} L. Th\'{e}venaz, Nat. Photonics 2, 474 (2008).
\bibitem{11} A. Shumelyuk, K. Shcherbin, S. Odoulov, B. Sturman, E. Podivilov and K. Buse, Phys. Rev. Lett. 93, 243604 (2004).
\bibitem{12} E. Baldit, K. Bencheikh, P. Monnier, J. A. Levenson and V. Rouget, Phys. Rev. Lett. 95, 143601 (2005).
\bibitem{13} S. Melle, O. G. Calder¨®n, C. E. Caro, E. Cabrera-Granado, M. A.
Ant¨®n and F. Carre$\tilde{n}$o, Opt. Lett. 33, 827 (2008).
\bibitem{14} K. Y. Song, M. G. Herr¨¢ez, L. Th¨¦venaz, Opt. Express 13, 9758 (2005).

\bibitem{15} T. J. Kippenberg and K. J. Vahala,  Science 321, 1172 (2008).
\bibitem{16} A. Schliesser and T. J. Kippenberg, arXiv: 1003. 5922v
(2010).
\bibitem{17} C. H. Metzger and K. Karrai, Nature 432, 1002 (2004).
\bibitem{18} O. Arcizet, P. F. Cohadon, T. Briant, M. Pinard and A. Heidmann, Nature 444, 71 (2006).
\bibitem{19} A. Schliesser, R. Rivi¨¨re, G. Anetsberger, O. Arcizet and T. J. Kippenberg, Nature Physics 4, 415 (2008).
\bibitem{20} W. He, J.J. Li and K. D. Zhu, Opt. Lett. 35, 339
(2010).
\bibitem{21} G. S. Agarwal and S. Huang, Phys. Rev. A 81, 041803
(2010).
\bibitem{22} T.J. Kippenberg, H. Rokhsari, T. Carmon, A. Scherer, and K. J. Vahala, Phys. Rev. Lett.  95, 033901 (2005).
\bibitem{23} A. Schliesser, P. Del'Haye, N. Nooshi, K. J. Vahala, and T. J. Kippenberg, Phys. Rev. Lett. 97, 243905 (2006).
\bibitem{24} V. Giovannetti and D. Vitali, Phys. Rev. A 63, 023812 (2001).
\bibitem{25} C. Genes, D. Vitali, P. Tombesi, S. Gigan, and M. Aspelmeyer, Phys. Rev. A 77, 033804 (2008).
\bibitem{26} M. Pinard, Y. Hadjar, and A. Heidmann, Eur. Phys. J. D 7, 107 (1999).
\bibitem{27} J. F. Lam, S. R. Forrest and G. L. Tangonan, Phys. Rev. Lett. 66, 1614 (1991).
\bibitem{28} B. I. Greene, J. F. Mueller, J. Orenstein, D. H. Rapkine, S. Schmitt-Rink and M. Thakur, Phys. Rev. Lett. 61, 325 (1988).
\bibitem{29} G. S. Agarwal and R. W. Boyd, Phys. Rev. A 60, R2681 (1999).
\bibitem{30} R. W. Boyd, Nonlinear Optics (Academic Press, Amsterdam) p.313 (2008).
\bibitem{31} D. F. Walls and G. J. Milburn, Quantum Optics (Springer-Verlag, Berlin, 1998), p. 124
\bibitem{32} S. Grolacher, K. Hammerer, M. R. Vanner and M. Aspelmeyer, Nature 460, 724 (2009).
\bibitem{33} S. E. Harris, J. E. Field and A. Kasapi, Phys. Rev. A 46, R29 (1992).
\bibitem{34} R. S. Bennink, R. W. Boyd, C. R. Stroud and V. Wong, Phys. Rev. A 63, 033804 (2001).

\end{thebibliography}
\end{document}